\documentclass{Interspeech2024}

\usepackage{multirow}
\usepackage{arydshln}

\interspeechcameraready

\title{An Investigation of Noise Robustness for Flow-Matching-Based Zero-Shot TTS}

\name[]{Xiaofei}{Wang}
\name[]{Sefik Emre}{Eskimez}
\name[]{Manthan}{Thakker}
\name[]{Hemin}{Yang}
\name[]{Zirun}{Zhu}
\name[]{Min}{Tang}
\name[]{Yufei}{Xia}
\name[]{Jinzhu}{Li}
\name[]{Sheng}{Zhao}
\name[]{Jinyu}{Li}
\name[]{Naoyuki}{Kanda}

\address{Microsoft Corporation, USA}
\email{\{Xiaofei.Wang, Sefik.Eskimez, Manthan.Thakker, Naoyuki.Kanda\}@microsoft.com}

\keywords{zero-shot TTS, noise-robust TTS, conditional flow-matching, generative pre-training, multi-task fine-tuning}

\newcommand{\ra}[1]{\renewcommand{\arraystretch}{#1}}

\begin{document}

\maketitle
 
\begin{abstract}
Recently, zero-shot text-to-speech (TTS) systems, capable of synthesizing any speaker’s voice from a short audio prompt, have made rapid advancements. However, the quality of the generated speech significantly deteriorates when the audio prompt contains noise, and limited research has been conducted to address this issue. In this paper, we explored various strategies to enhance the quality of audio generated from noisy audio prompts within the context of flow-matching-based zero-shot TTS. Our investigation includes comprehensive training strategies: unsupervised pre-training with masked speech denoising, multi-speaker detection and DNSMOS-based data filtering on the pre-training data, and fine-tuning with random noise mixing. The results of our experiments demonstrate significant improvements in intelligibility, speaker similarity, and overall audio quality compared to the approach of applying speech enhancement to the audio prompt.
\end{abstract}

\section{Introduction}
In recent years, text-to-speech (TTS) technology has made significant advancements~\cite{taylor2009text, tan2021survey, zhang2023survey}, achieving a level of naturalness comparable to human speech~\cite{tan2024naturalspeech}. 
Further advancements have been made towards a zero-shot TTS system~\cite{wang2023neural, borsos2023soundstorm, shen2023naturalspeech, jiang2023mega, le2023voicebox, wang2023speechx, gao2023e3, vyas2023audiobox, kanda2024making} that can generate any speaker's voice with minimal enrolled recordings, or an audio prompt. 
Zero-shot TTS has a wide range of applications, including speech-to-speech translation, audio/video content creation, and personal assistant services. 
However, one of the challenges faced by such systems is handling noise in the audio prompt. Existing zero-shot TTS models tend to generate speech with a style of noise similar to that contained in the audio prompt. 
This property is undesirable for many applications that require clean speech.
In this paper, we aim to develop a zero-shot TTS system that can generate high-quality clean speech from any speaker, regardless of the existence of background noise in the audio prompt.
We refer to this property as the noise robustness of zero-shot TTS.

While there has been a surge of research interest in zero-shot TTS technology, research on noise robustness is limited. 
The most naive approach involves applying speech enhancement (SE) to the audio prompt before feeding it to a zero-shot TTS model. While this approach is simple, even the latest SE models inevitably cause processing artifacts (e.g., \cite{iwamoto22_interspeech, eskimez21b_interspeech}), which result in degraded speech quality from the zero-shot TTS model. 
Our preliminary experiment revealed that the application of SE causes degradation in both intelligibility and speaker characteristics of the generated audio. 
Recently, Fujita et al.~\cite{fujita2024noise} proposed enhancing the noise robustness of zero-shot TTS by training a noise-robust speaker embedding extractor using a self-supervised learning model. 
While the authors reported promising results, their method is only applicable to a zero-shot TTS system based on speaker embeddings. 
Most recent zero-shot TTS models utilize in-context learning, such as neural-codec-based language modeling~\cite{wang2023neural, borsos2023soundstorm, wang2023speechx} or audio infilling~\cite{le2023voicebox, vyas2023audiobox, kanda2024making}, instead of representing the audio prompt as a speaker embedding. 
It is essential to study the noise robustness of zero-shot TTS in state-of-the-art model architectures.

In the spirit of advancing state-of-the-art technology, this paper presents our efforts to improve the noise robustness of flow-matching-based zero-shot TTS~\cite{le2023voicebox},
one of the leading models in terms of intelligibility and speaker characteristics preservation. 
We explored a range of training strategies, including generative pre-training~\cite{liu2023generative} with masked speech denoising, multi-speaker detection and DNSMOS~\cite{reddy2021dnsmos}-based data filtering on the pre-training data, as well as the fine-tuning with random noise mixing. 
Through experiments with both clean and noisy audio prompt settings, we demonstrate that intelligibility, speaker similarity, and overall audio quality can be consistently improved compared to an approach that applies SE to the audio prompt.
In addition, as a byproduct, we demonstrate for the first time that our zero-shot TTS model achieves better speaker similarity compared to the ground-truth audio in the widely used cross-utterance evaluation setting on LibriSpeech~\cite{panayotov2015librispeech}.

\section{Flow-matching based zero-shot TTS}
\subsection{Overview}
Our TTS system closely follows Voicebox~\cite{le2023voicebox},
which consists of a flow-matching-based audio model
and a regression-based duration model.
This section covers the overview of each model.

The objective of the audio model is to generate a log mel spectrum $\tilde{x} \in \mathbb{R}^{D\times T}$ given a frame-wise phoneme index sequence $a \in \mathbb{Z}_+^{T}$ under the condition that the value of $\tilde{x}$ is partially known as $x_{\text{ctx}} \in \mathbb{R}^{D \times T}$. Here, $D$ represents the feature dimension, and $T$ is the sequence length. 
$x_{\text{ctx}}$ is also known as the audio context, 
and the known value of $\tilde{x}$ is filled in; otherwise, the value is set to zero.
In the inference, $\tilde{x}$ is generated 
based on $x_{\text{ctx}}$ and $a$ where a part of $x_{\text{ctx}}$ is 
filled by the log mel spectrum of the audio prompt.
Based on the in-context learning capability of the model,
the speaker characteristics of the generated part of $\tilde{x}$ 
becomes similar to that of the audio prompt. 
The estimated $\tilde{x}$ is then converted to the speech signal based on a vocoder.

The audio model needs to be
 trained to enable sampling from $P(\tilde{x}|a, x_{\text{ctx}})$.
It is achieved 
 based on the flow-matching framework. 
This technique morphs a simple initial distribution $p_0$ into a more complex distribution $p_1$ that closely matches the observed data distribution. 
The model is trained based on the conditional flow-matching objective \cite{lipman2022flow}.
Specifically,
the model is trained to estimate
a time-dependent vector field $v_t, t\in[0,1]$, which is used to construct a flow $\phi_t$ that pushes the initial distribution towards the target distribution. 
The sampling process of $\tilde{x}$ is achieved by solving
the ordinary differential equation 
with the estimated vector field $v_t$ and 
initial random value sampled from $p_0$.
Refer \cite{lipman2022flow} for more details.

The duration model follows the regression-based approach detailed in~\cite{le2023voicebox}. 
This model takes a phoneme sequence $p \in \mathbb{Z}_+^{N}$,
where $N$ represents the number of phonemes. 
The model is trained to
predict the duration for each phoneme $\tilde{l} \in \mathbb{R}_+^{N}$
under the condition that the value of $\tilde{l}$ is partially known as
$l_{\text{ctx}} \in \mathbb{Z}_+^{N}$. 
Similar to the audio model, $l_{\text{ctx}}$ is filled by the known value of $\tilde{l}$, and the unknown part is filled by zero.
The model is trained based on the mean square error loss
on the predicted duration.
Refer \cite{le2023voicebox} for more details.

\subsection{Unsupervised pre-training of audio model}

Liu et al.~\cite{liu2023generative} proposed to pre-train the flow-matching-based audio model with a large amount of unlabeled training data. They reported superior audio model quality after fine-tuning. During pre-training, the phoneme sequence $a$ is dropped, and the model is trained to predict the distribution of $P(\tilde{x}|x_{\text{ctx}})$. For each training sample, $n$ non-consecutive random segments are selected
with a constraint of the minimum number of frames, $\texttt{Min}_F$, of 
each masked segment. In this work, we set $\texttt{Min}_F=5$ for all our exploration
based on our preliminary experiment.

\section{Approach toward noise robustness}
This section explains our approaches to enhance noise robustness through pre-training and fine-tuning of the flow-matching-based audio model. 

\subsection{Data filtering in pre-training}

We want to utilize a large amount of unlabeled data for pre-training to further improve the performance of the audio models. However, real-world data is often low-quality and noisy, and using such data without proper filtering can negatively impact the model performance. Therefore, to ensure the quality of our models, 
we explore data filtering techniques that can effectively identify and prioritize high-quality, noise-free samples for pre-training. Consequently, we employ the following two strategies to filter the pre-training data. 

Our first strategy involves filtering out the samples with more than one speaker. 
To detect the multiple speakers in an audio sample, 
we employ an in-house speaker change detection model, and discard a sample whenever the speaker change is detected.
Our second strategy involves assessing the speech quality of the samples. We employ the DNSMOS~\cite{reddy2021dnsmos}, a neural network-based mean opinion score estimator\footnote{\tiny\url{https://github.com/microsoft/DNS-Challenge/tree/master/DNSMOS}}, to evaluate the speech quality. We then discard samples that fall below a certain DNSMOS value threshold $\texttt{DNSMOS}_T$. In the experiments section, we explore the impact of different threshold values for our second strategy.

\subsection{Masked speech denoising in pre-training}
Masked speech denoising, introduced in WavLM~\cite{chen2022wavlm}, is an approach to enhance the model's ability to focus on relevant speech signals amid noise. It involves estimating clean audio for the masked part from the noisy audio input. Inspired by the success of WavLM, we investigate a similar approach for flow-matching-based model pre-training. 

During pre-training, 
in a probability of $P_\text{n}^{\text{pre}}$,
we simulate noisy speech by mixing training samples with randomly selected noise, which yields pairs of noisy speech and clean speech.
We use the noisy speech to extract the context input $x_{\text{ctx}}$, and the original training sample as the training target.
In the noise mixing phase, we randomly sample the noise from the DNS challenge corpus \cite{dubey2023icassp}, crop it, and 
mixed it with the signal-to-noise ratio (SNR) ranging from 0dB to 20 dB. 
We ensure that the duration of the noise does not exceed 50\% of that of the training audio.
We also explore the mixing of a secondary speaker into the audio with a probability $P_\text{s}^{\text{pre}}$, drawing parallels to WavLM.
The secondary speaker is picked from the same training batch of the primary speaker. All the mixing settings are the same as the noise mixing one, except that the SNR ranges between $[0, 10]$ dB.

\subsection{Fine-tuning with random noise mixing}
\label{sec:fine-tuning}
We also explore the fine-tuning strategy of the audio model. 
Conventionally, the audio model is fine-tuned with clean training data~\cite{liu2023generative}.
On the other hand, Fujita et al.~\cite{fujita2024noise} 
concurrently\footnote{The paper~\cite{fujita2024noise} was published on Jan 10th, 2024 when we were preparing our paper.} proposed to fine-tune
their zero-shot TTS model by including noise to the audio prompt 
in a 50\% ratio to improve the noise robustness.
In our work, we also explore 
the similar approach in the context of flow-matching-based zero-shot TTS.
Specifically,
we randomly add noise
in a probability $P_\text{n}^{\text{ft}}$ to the audio to extract 
the audio context $x_{\text{ctx}}$,
while the training target remains the original clean audio.
Noise samples from the DNS challenge corpus~\cite{dubey2023icassp} 
are randomly selected and mixed 
at SNRs between -5 dB and 20 dB.

\begin{table*}[t]
    \caption{Results on zero-shot TTS for cross-utterance settings with LibriSpeech test-clean. 
SE: speech enhancement, DF: data filtering, Hu: HuBERT-L, Ne: NeMo, Wa: WavLM}
\vspace{-3mm}
  \label{tab:main_results}
  \ra{0.9}
  \tabcolsep = 0.9mm
  \centering
 \resizebox{\textwidth}{!}{
 \footnotesize
\begin{tabular}{clccccccccccccccccc}
  \toprule
  &
\multicolumn{5}{c}{Zero-shot TTS system}  && 
\multicolumn{5}{c}{Clean prompt} && 
\multicolumn{5}{c}{Noisy prompt} 
\\ 
\cmidrule{1-6}
\cmidrule{8-12}
\cmidrule{14-18}
 & Model & SE & DF & $P_\text{n}^{\text{pre}}$ & $P_\text{n}^{\text{ft}}$ 
  & & 
WER (\%)$\downarrow$ && SIM-o$\uparrow$ && \multirow{2}{*}{DNSMOS$\uparrow$} && 
WER (\%)$\downarrow$ && SIM-o$\uparrow$ && \multirow{2}{*}{DNSMOS$\uparrow$} 
\\ 
& & & & & 
 &  & 
Avg. (Hu / Ne)  && Avg. (Wa / Ne)  &&  && 
Avg. (Hu / Ne)  && Avg. (Wa / Ne)  &&  
\\ 
\midrule
 &
 Ground truth &- &- &- &-    &&   
1.9 (2.1 / 1.7) && 0.74 (0.71 / 0.77) && 3.30 &&  
4.2 (5.1 / 3.3) && 0.71 (0.68 / 0.73) && 2.57
\\
 & Ground truth & $\checkmark$ &- &- &-    &&   
2.0 (2.1 / 1.8) && 0.74 (0.71 / 0.76) && 3.36 &&  
3.6 (3.8 / 3.3) && 0.68 (0.67 / 0.68) && 3.24    
\\
\midrule
& VALL-E~\cite{wang2023neural} &- &- &- &- && 
- (5.9 / -) && - && -  &&
- && - && - 
\\
& Naturalspeech 2~\cite{shen2023naturalspeech} &- &- &- &-&& 
- (2.3 / -) && - (0.62 / -) && -  &&
- && -  && -  
\\
& Voicebox~\cite{le2023voicebox} &- &- &- &-&& 
- ({\bf 1.9} / -) &&  - (0.66 / -) && -  &&  
-  &&     - && -  
\\
& SpeechFlow~\cite{liu2023generative} &- &- &- &- && 
- (2.1 / -)&& - (0.70 / -) && -  &&  
-  && -    && - 
\\ 
\midrule
(B1) & Our TTS model &- &- &- &-&&
2.7 (2.3 / {\bf 3.0}) &&  0.71 (0.67 / 0.75) && 3.33  &&
{\bf 2.5} (2.3 / {\bf 2.6})  && 0.60 (0.56 / 0.63)     && 3.01
\\
(B2) & Our TTS model & $\checkmark$ &- &- &-&&
2.8 (2.3 / 3.2) && 0.69 (0.66 / 0.72) && 3.37  && 
2.8 (2.3 / 3.3) && 0.60 (0.58 / 0.61) && 3.28
\\
\midrule
(P1) & Our TTS model & - & $\checkmark$ & - &-&&
{\bf 2.6} (2.2 / {\bf 3.0}) && {\bf 0.75} ({\bf 0.72} / {\bf 0.78}) && 3.35  &&
2.6 (2.4 / 2.7)  && {\bf 0.65} (0.62 / {\bf 0.68}) && 2.99 
\\
(P2) & Our TTS model & $\checkmark$ & $\checkmark$ & - &-&&
2.8 (2.3 / 3.2) && 0.74 (0.71 / 0.76) && {\bf 3.39} && 
2.9 (2.4 / 3.3) && 0.64 ({\bf 0.63} / 0.65) && 3.29 
\\
\hdashline[1pt/2pt]\hdashline[0pt/1pt]  
(P3) & Our TTS model & - & $\checkmark$ & - & 1.0&&
2.8 (2.3 / 3.2) && 0.74 (0.70 / 0.77) && 3.35  &&
2.7 (2.3 / 3.1) && 0.64 (0.61 / 0.67) && {\bf 3.32} 
\\
(P4) &   Our TTS model &- & $\checkmark$ & - &0.5&& 
2.7 (2.2 / 3.1) && 0.74 (0.71 / 0.77) && 3.34  &&
2.7 (2.3 / 3.1) && 0.64 (0.61 / 0.66) && 3.31 
\\
(P5) & Our TTS model & - & $\checkmark$ & 0.5  &0.5 &&  
{\bf 2.6} (2.2 / {\bf 3.0})  && {\bf 0.75} (0.71 / {\bf 0.78}) && 3.35  &&
2.6 ({\bf 2.2} / 2.9)  && {\bf 0.65} (0.62 / 0.67)  && {\bf 3.32}
\\
\bottomrule 
\end{tabular}
}
 \vspace{-3mm}
\end{table*}

\section{Experimental results}

\subsection{Training data}

The pre-training data of the audio model consisted of 200,000 hours of in-house unlabeled anonymized English audio, without undergoing any form of preprocessing. 
The audio occasionally included background noise, with significant variations in quality.

For fine-tuning of the audio and duration model, we used the LibriLight~\cite{kahn2020libri}, which consists of approximately 60,000 hours of untranscribed English audio from over 7,000 speakers~\cite{kahn2020libri}. 
Since LibriLight does not provide reference transcriptions, 
we transcribed the audio based on the off-the-shelf Kaldi automatic speech recognition (ASR)\footnote{\tiny\url{https://kaldi-asr.org/models/m13}}, and used the phoneme sequences from the ASR hypothesis to fine-tune the audio model, as similar to~\cite{le2023voicebox}.

\subsection{Training and inference configurations}
In our experiment, the architecture of the audio model closely followed the configurations in \cite{le2023voicebox}. 
Specifically, we used Transformer with 24 layers, features 16 attention heads, and an embedding dimension of 1024. It also included a feed-forward layer dimension of 4096, alongside 1024 dimensions for phone embeddings. The model underwent pre-training over 25.6 million iterations.
Linear-decay learning rate (lr) schedulers were used for both pre-training and fine-tuning, with a warm-up having 1/10 of the total number of updates and a peak lr at 7.5e-5.
Other investigations of hyperparameters will be discussed in the next session.
During the inference, we used classifier-guidance-free with a guidance strength of 1.0, and the number of function evaluations (NFE) was 32.
A BigVGAN~\cite{lee2022bigvgan}-based vocoder was used to convert the mel spectrum into waveforms.

As for the duration model, 
we used a regression-based masked duration model by closely following the configuration in~\cite{vyas2023audiobox}. Specifically, we used the following configuration: 8 layers, 8 attention heads, 512 embedding, and 2048 feed-forward dimensions. 
The model was trained with an effective mini-batch size of 120K frames for 600K mini-batch updates. 
We closely followed the training parameters in~\cite{vyas2023audiobox}.

\subsection{Evaluation data}
In this study, we designed two test settings to evaluate the performance of zero-shot TTS models: a clean-prompt setting and a noisy prompt setting.

{\bf Clean-prompt setting}:
To assess the zero-shot TTS capabilities of our model under neutral speech conditions, we conducted evaluations using the `test-clean' subset from the LibriSpeech dataset~\cite{panayotov2015librispeech}. 
Adhering to the prior works~\cite{le2023voicebox, wang2023neural, shen2023naturalspeech}, we selected audio samples with durations ranging from 4 to 10 seconds. For each sample, zero-shot TTS was performed using the transcription of the sample as a text prompt, and a 3-second audio clip from another randomly selected audio of the same speaker as an audio prompt.
Following \cite{le2023voicebox}, we used the final 3 seconds
of the randomly selected audio as the audio prompt.

{\bf Noisy-prompt setting}:
We prepared each test sample from the clean-prompt set by blending its audio prompt with a noise sample. The noise sample was randomly selected from the MUSAN dataset~\cite{snyder2015musan}. The signal-to-noise ratio (SNR) for the mixture was determined randomly, falling within a range of 0 dB to 20 dB. We then trimmed the final 3 seconds of the mixed sample and used it as the audio prompts for zero-shot TTS. 
The audio prompt selection mirrored the clean-prompt setting, ensuring the only difference between clean and noisy settings was the presence of noise in the audio prompt.

\subsection{Evaluation metrics}
We evaluated the generated speech based on the following metrics.
Note that we synthesized the audio with three random seeds and reported the average of them for all our experiments.

\textbf{Word error rate (WER):} We used the WER as a metric to evaluate the intelligibility of the generated audio. For our experiments, we employed two ASR systems: 
hubert-large-ls960-ft model\footnote{\tiny\url{https://huggingface.co/facebook/hubert-large-ls960-ft}}, which was used in most prior publications, and NeMo's stt\_en\_conformer\_transducer\_large model\footnote{\tiny\url{https://huggingface.co/nvidia/stt\_en\_conformer\_transducer\_xlarge}}, known for its superior stability and robustness against noise and processing artifacts.  
We report both the average WER and the individual WERs 
to credibly assess intelligibility.

\textbf{Speaker similarity score (SIM-o):} A speaker verification model was used to evaluate how closely the generated speech resembles the voice characteristics of the audio prompt.
Specifically, SIM-o was computed as the cosine similarity between the speaker embeddings between the generated speech and the original clean audio prompt, for both clean-prompt and noisy-prompt settings. 
We utilized two speaker embedding extraction models--the WavLM Large\footnote{\tiny\url{https://github.com/microsoft/UniSpeech/tree/main/downstreams/speaker\_verification}}, which was used in most prior works, and the NeMo's TitaNet-Large\footnote{\tiny\url{https://huggingface.co/nvidia/speakerverification\_en\_titanet\_large}} as another speaker verification model. 
We report both the average SIM-o and the individual SIMs 
to credibly assess the speaker similarity.

\textbf{DNSMOS:} For measuring the cleanness of the generated audio, we utilized the DNSMOS score. Specifically, we employed the OVRL score from the DNSMOS P.835 model~\cite{reddy2021dnsmos}.

\subsection{Result}
\subsubsection{Pre-analysis on the ground-truth audio}
The first row of Table.~\ref{tab:main_results} presents 
WER, SIM-o, and DNSMOS
of the ground-truth audio.
Unsurprisingly,
the ground-truth audio 
in the clean-prompt condition 
provided us with the low WER, high SIM-o, and high DNSMOS scores, and the addition of the noise
significantly deteriorate all metrics.
We employed the AlignCruse~\cite{indenbom2022deep} model, which was trained for both SE and personalized SE in a multi-task setting, using the training configuration described in~\cite{eskimez23_interspeech}. This hybrid model operates in SE mode when the speaker embedding is a zero vector and in personalized SE mode when the speaker embedding is a non-zero vector. We applied this model in SE mode to our noisy prompts to remove background noise.
As expected, the SE significantly improved
the DNSMOS score in the noisy-prompt condition.
However, it came with the cost of the notable degradation 
on the SIM-o in the noisy-prompt condition,
and also the slight degradation of WER in the clean-prompt 
condition.
This result
demonstrated the inherent difficulty
to improve all metrics
in both clean and noisy conditions.

\begin{table}[t]
  \centering
  \caption{Impact of pre-training data filtering and fine-tuning steps. MS: multi-speaker detection-based filtering.
  }
\vspace{-3mm}
    \ra{0.9}
  \tabcolsep = 0.9mm
  \centering
  \label{tab:ablation_lr}
\resizebox{0.48\textwidth}{!}{
  \begin{tabular}{@{}cccccccccccc@{}}
  \toprule
   \multicolumn{2}{c}{Pre-training} && Fine-tuning && \multicolumn{2}{c}{Clean Prompt}  && \multicolumn{2}{c}{Noisy Prompt} \\ 
   \cmidrule{1-2} \cmidrule{4-4} \cmidrule{6-7} \cmidrule{9-10}
   $\texttt{DNSMOS}_T$ & MS && Steps && WER (\%)$\downarrow$ & SIM-o$\uparrow$ && WER (\%)$\downarrow$ & SIM-o$\uparrow$\\ 
  \midrule
  0.0 & -  && 0.64M && 2.65 & 0.707 && 2.46 & 0.593 \\
  0.0 & $\checkmark$ && 0.64M && 2.66 & 0.719 && 2.48 & 0.612 \\
  2.6 & $\checkmark$ && 0.64M && 2.72 & 0.746 && 2.51 & 0.641 \\
  2.8 & $\checkmark$ && 0.64M && {\bf 2.62} & {\bf 0.749} && 2.51 & {\bf 0.647}  \\
  3.0 & $\checkmark$ && 0.64M && 2.64 & 0.746 && {\bf 2.43} & 0.638 \\
  \midrule
  \multirow{4}{*}{2.8}  & $\checkmark$ && 0.16M && 2.70 & {\bf 0.756} && 2.62 & {\bf 0.659} \\
                        & $\checkmark$ && 0.32M && 2.64 & 0.752 && 2.54 & 0.650 \\
                        & $\checkmark$ && 0.64M && {\bf 2.62} & 0.749 && {\bf 2.51} & 0.647 \\
                        & $\checkmark$ && 1.6M  && 2.64 & 0.745 && 2.98 & 0.633 \\
  \bottomrule 
\end{tabular}
}
 \vspace{-3mm}
\end{table}

\subsubsection{Main results}

The results of various zero-shot TTS models are presented from the 3rd row to the last row of Table.~\ref{tab:main_results}. Firstly, our baseline TTS model (model B1) showed decent WER and SIM-o compared to prior works~\cite{wang2023neural,shen2023naturalspeech,le2023voicebox,liu2023generative} in the clean prompt setting.\footnote{We did not list Audiobox~\cite{vyas2023audiobox} due to the difference of the experimental configuration.} However, the SIM-o and DNSMOS significantly dropped in the noisy prompt setting, indicating that the noise in the audio prompt was transferred into the generated speech. 
Unexpectedly, we observed an improvement in the WER from NeMo ASR in the noisy prompt, which resulted in a minor average WER improvement. This might be because NeMo ASR is sensitive to speech artifacts rather than noise, and small amounts of noise might conceal the artifacts within the speech portion.

Secondly, when we applied SE on the audio prompt (model B2), we observed a significant improvement in the DNSMOS score (3.01$\rightarrow$3.28) in the noisy prompt setting. However, it came with noticeable degradation of SIM-o in the clean prompt setting (0.71$\rightarrow$0.69) and degradation of WER in both clean (2.7\%$\rightarrow$2.8\%) and noisy prompt settings (2.5\%$\rightarrow$2.8\%).

We then applied our proposed pre-training data filtering based on multi-talker detection and DNSMOS-based filtering with $\texttt{DNSMOS}_T=2.8$ (model P1). It significantly improved WER and SIM-o for both clean and noisy prompt settings. 
However, in this condition, the application of SE on the audio prompt (model P2) still introduced significant degradation of WER and SIM-o as a trade-off for the DNSMOS improvement.

Next, we investigated the random application of noise in the fine-tuning as described in \ref{sec:fine-tuning}, where we set $P_\text{n}^{\text{ft}}=1.0$ (model P3) or $P_\text{n}^{\text{ft}}=0.5$ (model P4). Compared to model P1, both models P3 and P4 achieved a significant improvement in the DNSMOS score (2.99$\rightarrow$3.32 or 3.31), while showing a marginal degradation of WER and SIM-o.

We then applied noise in the pre-training with $P_\text{n}^{\text{pre}}=0.5$ (model P5), which resulted in small but consistent improvements in all metrics in both the clean and noisy prompts. Compared to our baseline model with SE (model B2), the final model P5 showed significant improvement in WER and SIM-o while keeping the DNSMOS score similarly high. We conducted paired t-tests on the metric values from models B2 and P5, and confirmed that all improvements on WER and SIM-o were statistically significant, with p-values less than 0.05.

Finally,
unexpectedly, we observed that our SIM-o of 0.75 for the clean prompt setting is even higher than that of the ground truth (0.74). Upon examination, we found that some speakers in the LibriSpeech use different voice characteristics to represent different characters in the book. This sometimes resulted in a low SIM-o for the ground truth in cross-utterance settings. To the best of our knowledge, this is the first time that a zero-shot TTS model has achieved a better SIM-o in the cross-utterance evaluation setting with LibriSpeech test-clean.

\begin{table}[t]
  \centering
  \caption{Impact of pre-training hyper parameters. 
  In this experiment, we pre-trained the TTS model with only 1.6M steps for rapid exploration.} 
\vspace{-3mm}
  \label{tab:simu}
    \ra{0.9}
  \tabcolsep = 1.5mm
  \centering
{
\footnotesize
  \begin{tabular}{@{}cccccccc@{}}
  \toprule
   \multicolumn{2}{c}{Pre-training} && 
   \multicolumn{2}{c}{Clean Prompt}  && 
   \multicolumn{2}{c}{Noisy Prompt}
   \\
   \cmidrule{1-2}\cmidrule{4-5}\cmidrule{7-8}
   $P_\text{n}^{\text{pre}}$ & 
   $P_\text{s}^{\text{pre}}$ && 
   WER (\%)$\downarrow$ & SIM-o$\uparrow$ && 
   WER (\%)$\downarrow$ & SIM-o$\uparrow$ \\
  \midrule
    0.00  & 0.00  && 2.64 & 0.698 && 2.65 & 0.577  \\
    0.25 & 0.00  && 2.68 & 0.702 && 2.61 & 0.580  \\
    0.50  & 0.00  && 2.64 & {\bf 0.706} && {\bf 2.58} & {\bf 0.585} \\
    0.75 & 0.00  && 2.65 & 0.693 && 2.61 & 0.577   \\
    1.00  & 0.00  && 2.72 & 0.689 && 2.66 & 0.572  \\
   \hdashline[1pt/2pt]\hdashline[0pt/1pt] 
    0.25 & 0.25 && {\bf 2.61} & 0.696 && 2.62 & 0.579  \\
    0.50  & 0.25 && 2.67 & 0.694 && 2.62 & 0.571 \\
  \bottomrule 
\end{tabular}
}
 \vspace{-3mm}
\end{table}

\subsubsection{Ablation studies}

Table~\ref{tab:ablation_lr} shows the impact of pre-training data filtering
and fine-tuning steps.
We first found that the multi-speaker detection-based data filtering,
which discarded 23.3\% of the pre-training data,
significantly improved SIM-o for both clean and noisy prompt settings.
We also found that 
the $\texttt{DNSMOS}_T=2.8$ achieved the highest SIM-o for
both clean prompt and noisy prompt settings
while achieving the lowest WER for 
the clean prompt.
These results suggest that 
our pre-training data filtering effectively filtered out low-quality audio samples 
while
keeping the diverse speaker data as much as possible.
We also found the gradual degradation in SIM-o scores as the number of fine-tuning steps increased from 0.16M to 1.6M, as shown in the lower half of the table.
This suggests that 
the diverse speaker characteristics learned by
the large-scale unsupervised pre-training can be 
deteriorated by the 
excessive fine-tuning with less speaker
diversity. 

Table \ref{tab:simu} shows the result of further investigations on
the pre-training hyperparameters. 
In this experiment,
we pre-trained the TTS model with only 1.6M steps for
rapid exploration.
We first observed that 
the random noise mixing during the pre-training effectively
improved SIM-o score, and observed the best result with $P_\text{n}^{\text{pre}}=0.5$ .
We also investigated the mixing of a secondary speaker's speech with
$P_\text{s}^{\text{pre}}=0.25$.
However, we did not observe any noticeable improvement from this trial.
Overall, we observed that appropriately injecting noise during the pre-training stage could effectively improve the zero-shot TTS model, not only for the noisy prompt setting but also for the clean prompt setting.

\section{Conclusions}
This paper presented our efforts to enhance the noise robustness of flow-matching-based zero-shot TTS. Our investigation covered a range of strategies, encompassing unsupervised pre-training with DNSMOS-based data filtering and masked speech denoising, as well as multi-task fine-tuning with random noise mixing. The results of our experiments demonstrated significant improvements in intelligibility, speaker similarity, and overall audio quality compared to the approach of applying speech enhancement to the audio prompt.

\bibliographystyle{IEEEtran}
\bibliography{mybib}

\end{document}